\documentclass[preprint2]{aastex}

\usepackage{epsf}
\usepackage{graphics}

\newcommand{\kms}{km s$^{-1}$}
\newcommand{\cts}{cts s$^{-1}$}
\newcommand{\ax}{$\alpha_{\rm X}$}
\newcommand{\aox}{$\alpha_{\rm ox}$}
\newcommand{\aro}{$\alpha_{\rm ro}$}
\newcommand{\cm}{cm$^{-2}$}
\newcommand{\rb}[1]{\raisebox{1.5ex}[-1.5ex]{#1}}
\newcommand{\msun}{$M_{\odot}$}

\newcommand{\pl}{$\pm$}
\newcommand{\nh}{$N_{\rm H}$}
\newcommand{\etal}{{\it et al.}}
\shorttitle{RX J1028--0844 and BR 0351--1034}
\shortauthors{Grupe \etal}

\begin{document}

\def\clipfig#1{\def\lbracket{[}\def\testit{#1}%
    \ifx\testit\lbracket\let\next=\optclipfig\else\let\next=\stdclipfig\fi%
    \next{#1}}
%
\newcommand {\hclipfig} [7] {\clipfig[#7]{#1}{#2}{#3}{#4}{#5}{#6}}
%
\def\usemodepsfig {\global\def\cfmode{x}\typeout{*** set clipfig to PSFIG mode ***}}
\def\usemodeepsf  {\global\def\cfmode{}\typeout{*** set clipfig to EPSF mode ***}}
\def\useunitmm    {\global\def\cfunit{x}\typeout{*** set clipfig to use mm as unit ***}}
\def\useunitcm    {\global\def\cfunit{}\typeout{*** set clipfig to use cm as unit ***}}
\def\clipfigsettings {\ifx\cfmode\empty\def\ccfmode{EPSF }\else\def\ccfmode{PSFIG }\fi%
    \ifx\cfunit\empty\def\ccfunit{cm }\else\def\ccfunit{mm }\fi%
    \typeout{*** current clipfig settings: \ccfmode mode, using \ccfunit as unit ***}}
%
%
%
%
\def\stdclipfig#1#2#3#4#5#6{\ifx\cfmode\empty%
    \let\next=\eclipfig\else\let\next=\pclipfig\fi%
    \next{#1}{#2}{#3}{#4}{#5}{#6}}
\def\optclipfig#1#2]#3#4#5#6#7#8{\ifx\cfmode\empty%
    \let\next=\ehclipfig\else\let\next=\phclipfig\fi%
    \next{#3}{#4}{#5}{#6}{#7}{#8}{#2}}
%
%
%
\newcommand {\pclipfig}[6] {\ifx\cfunit\empty%
        \psfig{figure=#1.ps,width=#2cm,bbllx=#3cm,bblly=#4cm,bburx=#5cm,%
           bbury=#6cm,clip=}\else%
        \psfig{figure=#1.ps,width=#2mm,bbllx=#3mm,bblly=#4mm,bburx=#5mm,%
           bbury=#6mm,clip=}\fi}
\newcommand {\phclipfig}[7] {\ifx\cfunit\empty%
        \hspace{#7cm}\psfig{figure=#1.ps,width=#2cm,bbllx=#3cm,bblly=#4cm,%
           bburx=#5cm,bbury=#6cm,clip=}\else%
        \hspace{#7mm}\psfig{figure=#1.ps,width=#2mm,bbllx=#3mm,bblly=#4mm,%
           bburx=#5mm,bbury=#6mm,clip=}\fi}
%
%
%
\newcommand {\eclipfig}[6]{%
  \ifx\cfunit\empty\epsfxsize=#2cm\else\epsfxsize=#2mm\fi%
  \epsfclipon\epsfverbosetrue%
  \cfcmtopspts{#3}\cfllxi=\cftempi\cfllxf=\cftempf%
  \cfcmtopspts{#4}\cfllyi=\cftempi\cfllyf=\cftempf%
  \cfcmtopspts{#5}\cfurxi=\cftempi\cfurxf=\cftempf%
  \cfcmtopspts{#6}\cfuryi=\cftempi\cfuryf=\cftempf%
  \def\cfstra{\number\cfllxi.\number\cfllxf}%
  \def\cfstrb{\number\cfllyi.\number\cfllyf}%
  \def\cfstrc{\number\cfurxi.\number\cfurxf}%
  \def\cfstrd{\number\cfuryi.\number\cfuryf}%
  \hbox{\epsfbox[{\cfstra} {\cfstrb} {\cfstrc} {\cfstrd}]{#1.ps}}}
\newcommand {\ehclipfig}[7]{%
  \ifx\cfunit\empty\epsfxsize=#2cm\else\epsfxsize=#2mm\fi%
  \epsfclipon\epsfverbosetrue%
  \cfcmtopspts{#3}\cfllxi=\cftempi\cfllxf=\cftempf%
  \cfcmtopspts{#4}\cfllyi=\cftempi\cfllyf=\cftempf%
  \cfcmtopspts{#5}\cfurxi=\cftempi\cfurxf=\cftempf%
  \cfcmtopspts{#6}\cfuryi=\cftempi\cfuryf=\cftempf%
  \def\cfstra{\number\cfllxi.\number\cfllxf}%
  \def\cfstrb{\number\cfllyi.\number\cfllyf}%
  \def\cfstrc{\number\cfurxi.\number\cfurxf}%
  \def\cfstrd{\number\cfuryi.\number\cfuryf}%
  \ifx\cfunit\empty\hspace{#7cm}\else\hspace{#7mm}\fi%
  \hbox{\epsfbox[{\cfstra} {\cfstrb} {\cfstrc} {\cfstrd}]{#1.ps}}%
  \vspace{-1mm}}
%
%
%
\newdimen\cfllxi \newdimen\cfllyi  \newdimen\cfurxi  \newdimen\cfuryi
\newdimen\cfllxf \newdimen\cfllyf  \newdimen\cfurxf  \newdimen\cfuryf
\newdimen\cftemp \newdimen\cftempi \newdimen\cftempf
\newdimen\cfpspoint \cfpspoint=1bp
%
%
%
\newcommand{\cfcmtopspts}[1]{\ifx\cfunit\empty%
  \cftemp=#1cm\else\cftemp=#1mm\fi%
  \multiply\cftemp10\divide\cftemp\cfpspoint%
  \cftempf=\cftemp\divide\cftemp10\cftempi=\cftemp\multiply\cftemp10%
  \advance\cftempf-\cftemp}
%
%
\def\cfmode{}\def\cfunit{}\clipfigsettings
%

\useunitmm

\def \charthoffset {\hspace{0.2cm}} \def \charthsep {\hspace{0.3cm}}
\def \chartvsepcap {\vspace{0.3cm}}
\def \chartvsep {\vspace{0.1cm}}
\newcommand{\putchartb}[1]{\clipfig{#1}{90}{5}{10}{250}{180}}
\newcommand{\putchartc}[1]{\clipfig{#1}{80}{5}{3}{250}{190}}
\newcommand{\chartlineb}[2]{\parbox[t]{18cm}{\noindent\charthoffset\putchartb{#1}\charthsep\putchartc{#2}\chartvsep}}
\newcommand{\chartlinec}[2]{\parbox[t]{18cm}{\noindent\charthoffset\putchartc{#1}\charthsep\putchartc{#2}\chartvsep}}


\title{XMM-Newton Observations of two high redshift quasars: RX J1028--0844
and BR 0351--1034
\thanks{Based on observations with XMM-Newton, an ESA Science Mission with
instruments and contribution directly funded by ESA member states and the 
U.S.A.
(NASA).
}
}


\author{D. Grupe, S. Mathur}
\affil{Astronomy Department, Ohio State University,
    140 W. 18th Ave., Columbus, OH-43210, U.S.A.}

\email{dgrupe, smita@astronomy.ohio-state.edu}

\author{B. Wilkes, and M. Elvis}
\affil{Harvard-Smithsonian Center for Astrophysics, 60 Garden Street,
    Cambridge, MA 02138, U.S.A.}





\begin{abstract}
We report the results of XMM-Newton observations of two high
redshift quasars, one radio-loud, RX J1028.6--0844 (z=4.276) and one
radio-quiet, BR 0351--1034 (z=4.351).  We find that the evidence
for strong excess absorption towards RX J1028--0844 is marginal at
best, contrary to previous claims.  The superior sensitivity and 
broader, softer energy range of XMM-Newton (0.2-10 keV)
allows better determination of spectral parameters
than much deeper ASCA observations (0.8-7 keV). Our XMM-Newton
observations call into
question several other ASCA results of strong absorption towards high
redshift radio-loud quasars. 
RX J1028.6--0844 occupies the same
parameter space in broad band spectral properties as the low redshift
BL Lac objects, showing no obvious evolution with redshift. The
radio-quiet quasar BR 0351--1024 became fainter between ROSAT and XMM
observations by a factor of at least 5, but with the present data we
cannot determine whether there is an associated spectral change. These
observations do not support previous claims of weaker X-ray emission
from high redshift radio-quiet quasars. The soft X-ray spectral slope
required to reconstruct the ROSAT PSPC hardness ratio of BR 0351--1034
is about
\ax=3.5, the steepest X-ray slope ever observed in a high redshift quasar,
and similar to that of low redshift Narrow Line Seyfert 1 galaxies.
\end{abstract}

\keywords{galaxies: active - quasars:general - quasars: individual 
(RX J1028--0844, BR0351--1034)
}

\section{Introduction}

High redshift quasars are interesting not only for their record setting
quality but also because they can tell us about the formation of quasars
and about conditions in the first few percent of the age of the
Universe. They allow us to study the evolution of a quasar's
central engine e.g. \citep{vig01}, the star formation
in the early Universe (e.g. \citet{die02a}), and the intergalactic
medium between the high redshift quasar and us (e.g. \citet{per01}). Prior
to ROSAT \citep{tru83} only one quasar with z$>$ 4.0
was detected in X-rays, GB 1508+5714 (z=4.30, \citet{mat95}). Only one
high redshift quasar was discovered during the ROSAT All-Sky Survey
(RASS, \citet{vog98}), RX J1028.6--0844 \citep{zic97}. The first
X-ray selected high redshift quasar was RX J1759.4+6632 (z=4.320,
\citet{hen94}) found in a deep ROSAT Position Sensitive Proportional
Counter (PSPC, \citet{pfe86}) observation. Other sources were detected
in X-rays, but selected in other wavelength bands, typically by their radio
emission, e.g. GB 1428+4217 (z=4.72, \citet{bol00}) or at optical
wavelengths (e.g. Q0000--263, z=4.111, \citet{bec94}). Thanks to the
Sloan Digital Sky Survey (SDSS, \citet{yor00}) the number of high
redshift quasars, even to redshifts z$>$6, has increased dramatically
 and several of them have been detected in X-rays. (e.g.
\citet{mat02, bra02, vig03a})\footnote{A complete list of z$>$4
quasars with X-ray detections is given at
www.astro.psu.edu/users/niel/papers/highz-xray-detected.dat}.

While detection of these $z>4$ quasars in X-rays has opened up a new
field of research, the results are conflicting.  For example,
\citet{bri97} and \citet{bec01} find that high redshift quasars are
more X-ray quiet and have flatter X-ray spectra than the low z
quasars, in contradiction to the \citet{mat02} results.
\citet{vig03a} also advocated that high redshift quasars are more
X-ray weak, but from a larger sample \citet{vig03b} concluded that the
trend is with luminosity rather than redshift.  One main reason behind
these contradictory results is that they have all been based on short,
snapshot observations.  The resulting total counts, of order of a few
tens, are generally too few to permit spectral analysis.  As a result, 
derived quantities such as $\alpha_{\rm ox}$\footnote{The X-ray loudness 
is defined by \citep{tan79} as
\aox=--0.384 log($f_{\rm 2keV}/f_{2500\AA}$).} 
have a strong dependence on the
underlying assumptions of spectral shape and absorbing column
density. To understand high redshift quasars, and to compare them
to their low redshift cousins, we have initiated a program to obtain X-ray
spectra of high redshift quasars using XMM-Newton. The sample consists
of both radio-loud and radio-quiet quasars to probe differential
evolution between the two classes, if any. Here we present results of
the AO 1 observations of a radio-loud quasar RX J1028--0844 and a
radio-quiet quasar BR 0351$-$1034.

The high redshift quasar RX J1028.6--0844 (RASS position:
$\alpha_{2000}$=10h 28m 38.9s;
$\delta_{2000}~=~-08^{\circ}44^{'}29{''}$; z=4.276) was identified by
\citet{zic97} in an identification program of northern X-ray sources
\citep{app98} detected during the RASS. Its PSPC count rate during
the RASS was 0.035\pl0.011 \cts~ which transfers to a rest-frame 2-10
keV luminosity $L_{\rm 2-10 keV}~=~6.4~\times~10^{46}$ ergs s$^{-1}$
which makes it one of the most X-ray luminous sources in the Universe
\citep{zic97}. RX J1028.6--0844 is associated with a close by radio
source PKS B1026--084 with flux of 220 mJy at 5 GHz in the
Parkes radio survey \citep{otr91}. From its
extreme luminosities at all wavelengths and its radio loudness RX
J1028.6--084 is considered a BL Lac object \citep{yua00}. In a long
67 hour observation by ASCA \citep{tan94}, \citet{yua00} found
evidence for very high neutral absorption at the rest-frame of RX
J1028.6--0844. Assuming solar abundances an absorption column of
2$\times~10^{23}$\cm~was found. However, it was not clear from the
ASCA data whether this absorption is associated with the quasar or if
it is related to a damped Ly$\alpha$ absorber at z=3.42
\citep{per01}.

The high redshift radio quiet quasar BR 0351--1034
($\alpha_{2000}$=03h 53m 46.9s,
$\delta_{2000}~=~-10^{\circ}~25^{'}~19.0^{''}$, z=4.351) was
discovered by the APM high-redshift quasar survey by
\citet{irw91}. \citet{sto96a} reported that BR0351--1034 was one of
the most unusual sources of their survey of high-redshift APM Quasars
with intervening absorption systems. They found saturated CIV 
absorption and a large number of absorption lines associated with
damped Lyman $\alpha$ absorption systems at z=3.633, 4.098, and 4.351.
The source was first detected in X-rays by ROSAT in a 9.1 ks pointed
PSPC observation with 54\pl13 counts \citep{kas00}.

In this paper we present the results of the XMM-Newton \citep{jan01}
observations of these two quasars. The short (5 ks) observation of RX
J1028.6--0844 was planned before the long ASCA observation.  The
supreme sensitivity of the EPIC PN detector \citep{str01} at soft
X-rays and recent calibration efforts, allow measurements down to 0.2
keV (or even less, \citet{hab03}), putting better constrains on the
intrinsic absorption of the source than from previous X-ray
missions. Our 26 ks observation of BR 0351--1034 was severely
affected by the high background radiation. As a result, the spectral
quality of the source was significantly compromised, and so the
spectral parameters are not well constrained.

The paper is organized as follows: in \S\,\ref{observe} we describe the
observations and data reduction, in \S\,\ref{results} we present the
results of the X-ray observation which will be discussed in
\S\,\ref{discuss}.

Throughout the paper spectral indeces are energy spectral indeces
with $F_{\nu} \propto \nu^{-\alpha}$. Luminosities are calculated
assuming a $\Lambda$CDM cosmology with $\Omega_{\rm M}$=0.3,
$\Omega_{\Lambda}$=0.7, and 
a Hubble constant of $H_0$ =75 \kms Mpc$^{-1}$, using the formulae given to
derive the luminosity distances given by  \citet{hogg99}. 
 All errors are 1$\sigma$ unless stated otherwise.

\section{\label{observe} Observations and Data Analysis}

\subsection{Radio-loud quasar RX J1028.6--0844}

RX J1028.6--0844 was observed by XMM-Newton on May 15th, 2002 for a total of
5 ks with the EPIC PN \citep{str01} and 7.3 ks with the EPIC MOS
\citep{tur01} detectors using thin filters. A high background flare
was present during a short time of the observation. We excluded this
time by creating a good time interval (GTI) file accepting only times when
the background count rate of photons with energies $>$ 10 keV was less
than 10 \cts. Only the EPIC PN observation was significantly affected
by the flare. The GTI screening of the PN data results in a total
observing time of 4350 s. Source photons in the EPIC PN were collected
in a circle with a radius of 35$^{''}$ and the background photons in a
circle of a radius of 60$^{''}$ close by. The source photons in the
MOS detectors were selected in a circle with a radius of 31.25$^{''}$
and the background from an annulus of 35$^{''}$ inner radius and
75$^{''}$ outer radius. We selected single and double events 
(PATTERN $\le$ 4) for the PN and single, double, triple and quadruple events
(PATTERN$\le$ 12) for the MOS, for which the
detectors are calibrated.

In addition to the XMM-Newton data of RX J1028.6--0844 we also retrieved the
ASCA data from observation of November 25th, 1999 (seq-id: 77011000;
\citet{yua00}) from the ASCA data archive at Goddard Space Flight
Center in order to fit our XMM-Newton data together with the ASCA
data. Due to the decrease in efficiency below 1 keV of the ASCA
Solid-state Imaging Spectrometers (SIS) after late 1994 (see ASCA
webpage heasarc.gsfc.nasa.gov/docs/asca/watchout.html) we
considered only photons with $E>0.8$ keV for our analysis.

\subsection{Radio-quiet quasar BR 0351--1034}

BR 0351--1034 was observed by XMM-Newton on 2002-08-23 for a total of 26.0 ks
with the EPIC PN and 27.7 ks with the EPIC MOS detectors using thin
filters. Due to a high background radiation during the last half of
the observation, these data were unusable. The data were screened to
create a GTI file with background count rate of photons with
energies $>$ 10 keV to be less than 10 \cts.  This screening results
in total observing times of 15.8 ks and 19.5 with the PN and MOS
detectors, respectively.

Source counts were collected in a circle with a radius of 15$^{''}$
and the background from a near by circular region with a radius of
30$^{''}$. Because of its small number of counts (Sect. \ref{res_xmm})
only the EPIC PN data with single and double events (PATTERN$\leq$4)
were used for the spectral analysis. Because of the faintness of the
source, the results may be affected by the choice of the background region.
We determined background from another near by region as well, and found
the results with the two different backgrounds to be consistent with
one another within errors.

The source was observed in a pointed ROSAT PSPC observation on
1992-01-27 (Obs-ID rp700531; \citet{kas00}) for a total of 9.4 ks. We
retrieved the data from the ROSAT Archive at MPE Garching and
reanalyzed them. The source photons were collected in a circle with
R=75$^{''}$. Since the source is faint, again we selected two
background regions, one in a circle with R=150$^{''}$ near the source
and a second one in an annulus around the source with an inner radius
R=100$^{''}$ and an outer radius R=200$^{''}$. Again, we found that
the choice of the background did not affect our results.

For both the sources, the XMM-Newton data were reduced using the XMM-Newton
Science Analysis Software (XMMSAS) version 5.3.3 and the X-ray spectra
were analyzed using XSPEC 11.2.0. The spectra were grouped by GRPPHA
3.0.0 in bins of at least 20 counts per bin. For both sources, the
Ancillary Response Matrix and
the Detector Response Matrix were created by the XMMSAS tasks {\it
arfgen} and {\it rmfgen}. The ASCA spectral data of RX J1028.6--0844 were
selected by XSELECT, grouped by GRPPHA like the XMM-Newton data, and analyzed by
XSPEC. 
The ROSAT data of BR 0351--1034 were analyzed  using the
Extended X-ray Scientific Analysis System (EXSAS, \citet{zim98})
version 01APR.  For the count-rate conversions between different X-ray
missions, PIMMS 3.2 was used.

\section{\label{results} Results}

\subsection{RX J1028.6--0844}

\subsubsection{\label{xresults} The XMM-Newton observation}

The mean count rates measured for the EPIC PN, MOS1 and MOS2 were
0.416\pl0.010 \cts, 0.115\pl0.004 \cts, and 0.094\pl0.005 \cts,
respectively. On the short time scale of the observation (the 5 ks
EPIC PN observation converts to $\approx$ 1ks in the object's
rest-frame) we could not detect any significant variability. We
converted the EPIC PN count rate into ROSAT PSPC and ASCA SIS count
rates by using PIMMS and could not detect any significant changes in
the count rate on long time scales either. None of the observations so far
have shown any significant variability in this
source. \citet{yua00} discussed the temporal analysis of their 67 h
ASCA observation and also found no significant variability.

Table\,\ref{rxj1028_spec} summarizes the results of a simple power law
fit with 'cold' absorption of neutral elements at z=0 to the
XMM-Newton data of RX J1028.6--0844. The fits to all instruments
separately agree within the errors, except for the MOS-1 which shows a
slightly higher absorption column than the other detectors. This
latter is driven by one data point at about 0.7 keV (see
Figure\,\ref{q1028_plot}), so the discrepancy does not appear to be
significant.  The X-ray slope is about \ax=0.3. This is in good
agreement with what has been found for other high redshift radio-loud
quasars (e.g. \citet{fer03}). When data from all the three XMM-Newton
detectors are fitted together with absorption fixed at the Galactic
value ($N_H~=~4.59~\times~10^{20}$\cm; \citet{dic90}), the fit is good
with $\chi^2=169$ for 160 degrees of freedom (DOF, see
Table\,\ref{rxj1028_spec} and Figure\,\ref{rxj1028_pn_wapo}). If the
absorbing column density is left as a free parameter, the fit yields
$N_H~=~7\pm1~\times~10^{20}$\cm~ and $\chi^2$=174.6 (161 DOF). Thus
the best fit $N_H~$ is consistent with the Galactic value, within
errors, and the spectra do not show any compelling evidence of excess
absorption.


\subsubsection{\label{asca_comp} Comparison with ASCA spectra}

The above result contradicts earlier ASCA results, reported by
\citet{yua00}, who found strong excess absorption towards the
source. To quantify the difference and to verify the ASCA results we
fitted the XMM-Newton and ASCA data with a power law plus galactic
absorption (fixed to the galactic value) and intrinsic redshifted
neutral absorption at the redshift of the quasar. We used the XSPEC
model {\it zvfeabs} for the redshifted intrinsic absorption. 
The metal abundance $Z/Z_{\odot}$ was set to unity (solar
abundance). The results of these fits are given in
Table\,\ref{rxj1028_spec_zvfeabs}. Here the ASCA SIS detectors show a
very high intrinsic column density of $N_{\rm H, intr}$ = 15.2\pl14.0$
\times 10^{22}$ cm$^{-2}$ which confirms the results found by
\citet{yua00}.  However, fits to the PN data result in much lower
column densities, as discussed above. Higher column densities are
obtained with MOS detectors, which do not have well calibrated
response below 0.5 keV and even higher column densities are found with
ASCA SIS which has no response below 0.8 keV. Thus the inferred high
column density appears to be a direct result of lack of low energy
response of these detectors.

To better understand the difference between XMM-Newton and ASCA results, we
performed the following additional tests. At first, we fitted the data by
fixing the intrinsic column density and X-ray slopes to the ones found
by ASCA ($N_{\rm H,intr}~=~21.1\times~10^{22}$\cm, \ax=0.72,
\citet{yua00}). In the 0.2-1.0 keV range we can clearly see strong
deviations at energies below 0.8 keV (Figure
\ref{rxj1028_pn_ascapara}). However, at energies above 0.8 keV the
data agree very well with the values found from the fits to the ASCA
data. Using the XMM-Newton and ASCA data only in the ASCA SIS 0.8-6.5
keV energy range results in much higher column densities than when the
whole detector energy ranges are used (Table
\ref{rxj1028_spec_zvfeabs}). The data in the 0.8-6.5 energy range of
the XMM-Newton detectors are indeed more or less consistent with the
parameters found by ASCA. Clearly, the lack of low energy response of
ASCA SIS gives incorrect results, and the superior sensitivity of
XMM-Newton in the soft energy band down to 0.2 keV allows us to make
accurate measurement of the absorbing column density.

Figure \ref{q1028_plot} displays the result of a powerlaw
fit with galactic and intrinsic absorption simultaneously to the
XMM-Newton PN and MOS-1 \& MOS-2 and ASCA SIS-0 \& SIS-1 data (Table
\ref{rxj1028_spec_zvfeabs}). All the data including the ASCA SIS data
are well-fitted with an X-ray spectral slope \ax=0.3 and an intrinsic
absorption column $N_{\rm H, intr}~\approx~10^{22}$ cm$^{-2}$ with
solar metal abundance, an order of magnitude lower than the column
density based on ASCA data alone.

To summarize, the XMM-Newton spectra do not show strong evidence for
excess absorption towards RX J1028-0844. If there is excess absorption
at all, and if it is at the redshift of the quasar, then it is an
order of magnitude lower than that inferred from ASCA data,


\subsubsection{Broad band Properties}

RX J1028.6--0844 is associated with a nearby radio source PKS
B1026--084 within the RASS error circle \citep{zic97} and references
therein; radio position: $\alpha_{2000}$=10h 28m 39.0s,
$\delta_{2000}=-08^{\circ} 44^{'} 36^{''}$ ).  The X-ray source
position is better determined with the XMM-Newton observation to be
$\alpha_{2000}$=10h 28m 38.9s, $\delta_{2000}=-08^{\circ} 44^{'}
39.2^{''}$ for the PN camera. Thus the difference in radio and X-ray
source positions is only $3.2^{''}$, well within the XMM-Newton
absolute pointing accuracy of $15^{''}$, and also within the boresight
error of $3^{''}-4^{''}$ \citep{ehle03}, making the
identification secure. 

 The radio flux at 5 GHz given in the Parkes Catalogue \citep{otr91}
 is 220 mJy. 
 This corresponds to a k-corrected
 luminosity density of log $l_{\rm 4.85 GHz}$=34.6 erg s$^{-1}$
 Hz$^{-1}$. The radio-to-optical slope $\alpha_{\rm ro}$\footnote{The
 radio-to-optical spectral slope is defined by \citep{zam81} as
 $\alpha_{\rm ro}~=~-$0.185 log$(l_{\rm 2500\AA}/l_{\rm 4.85~GHz}$)}
 was calculated from the k-corrected luminosity densities to be 
 $\alpha_{\rm ro}$ = 0.48. This converts to a radio loudness
 R\footnote{Radio loudness R = $f_{\rm 5~GHz}/f_{\rm 4400\AA}$}
 $\approx$180 which makes this source a radio-loud object as per
 the definition of
\citet{kel89}; radio-quiet, -loud division at R=10. The radio
 spectral slope is $\alpha_r=-0.3$ \citep{zic97}, making this a flat
 spectrum radio-loud object, and the high luminosity suggests a
 possible BL Lac identification \citep{yua00}. The slope of the
 X-ray power-law of the source is also flat (\ax$\approx$0.3;
 Table\,\ref{rxj1028_spec}), similar to other radio-loud sources (e.g
 \citep{elv86, bri97b}).

From the R magnitude R=17.1 mag given at the position of RX
J1028--0844 in the US Naval Observatory Catalogue A2.0 we estimated
the rest frame flux density at 2500\AA~ to log$f_{\rm 2500\AA}$ =
--26.4 ergs s$^{-1}$ cm$^{-2}$ Hz$^{-1}$ assuming an UV spectral
slope $\alpha_{\rm UV}$=0.8 as given in \citet{vig01, vig03a} and a
correction for de-reddening with $\frac{N_{\rm H}}{E_{\rm
B-V}}~=~4.93~10^{21}$\cm~mag$^{-1}$ \citep{dip94}.  The unabsorbed
rest-frame 2keV flux density is log $f_{\rm 2 keV}$ = $-29.0$ ergs
s$^{-1}$ cm$^{-2}$ Hz$^{-1}$, which results in an X-ray loudness of
\aox=1.08. This is different from
the value \aox=0.79 given in \citet{vig01}, but comparable to the
\aox~ values of other high redshift BL Lac sources. The difference in
the two values of \aox~ is due to the different values of R magnitude
used as inputs. We obtained the R magnitude from the USNO A2.0
catalogue while \citet{vig01} used an R magnitude from the literature,
probably R=18.9 given in \citet{zic97}.

\subsection{BR 0351--1034}

\subsubsection{\label{res_xmm} The XMM-Newton Observation}

The mean count rates are (6.26\pl2.39) 10$^{-3}$ \cts~ for the EPIC
PN, and (2.93\pl0.48) 10$^{-3}$ \cts~ and (2.29\pl0.48) 10$^{-3}$
\cts~ for the EPIC MOS 1 and 2, respectively. This results in total
numbers of background subtracted source counts of 100 for the PN and
58 and 45 for the MOS 1 and 2. The number of counts in the PN detector
are enough, while those in MOS 1 and MOS 2 are not, to perform a
simple spectral analysis. So we focus only on the PN in the following
for spectral analysis.

Table\,\ref{br0351_spec} summarizes the results of spectral fits to the
EPIC PN data of BR 0351--1034. The first fit was a simple powerlaw with
'cold' absorption of neutral elements in our Galaxy ($N_{\rm H,
gal}~=~4.08~\times~10^{20}$\cm; \citet{dic90}). The fit is good with
$\chi^2=2.3$ for 7 DOF. Figure\,\ref{br0351_specfits_galnh}
displays this fit to the PN data of BR 0351--1034.
If absorption is allowed
to be a free parameter, the best fit column density is
consistent with the Galactic column density
within errors. However, given the low data quality, we cannot rule out
intrinsic absorption of the order of a few times $10^{22}$\cm
(Table \ref{br0351_spec}, Figure \ref{br0351_speccontour}). Motivated by
the ROSAT result (see below), we also tried a fit with a broken
powerlaw model with a soft X-ray spectral slope $\alpha_{\rm
X,soft}$=3.5 as derived from the ROSAT data. The fit is good, implying
consistency with the ROSAT data but the data quality is
not high enough to provide a conclusive result on the presence of
a broken power-law.

\subsubsection{\label{br0351_rosat} Comparison with ROSAT}

The count rate during the ROSAT PSPC observation was determined using
both background regions in the PSPC; both rates agree within the errors
with a rate of 3.63\pl1.34 10$^{-3}$\cts. Based on the PSPC count rate,
we expect a PN count rate CR=0.039 \cts~ for a simple powerlaw model
with Galactic absorption (see Table \ref{br0351_spec}), but measured
only about 1/6th of it. Clearly the source varied in intensity between
the ROSAT and XMM-Newton observations.

The number of photons from the pointed ROSAT PSPC observation was
34.2\pl12.7, not sufficient for a spectral analysis. However, it is
possible to derive a hardness ratio\footnote{HR=(H-S)/(H+S) with S
counts in the 0.1-0.4 keV range and H=0.5-2.0 keV} in order to see
whether the spectrum is soft or hard. The measured hardness ratio is 
HR=--0.24\pl0.57 and implies a steep 
soft X-ray spectrum. In order to get
a hardness ratio HR=--0.24 for the PSPC data, a powerlaw model with
galactic absorption requires a spectral slope \ax=3.5. This is a very
steep X-ray slope and would be even steeper if the source is also intrinsically
absorbed. Thus it is unlikely that there was excess absorption towards
the source during the time of the
ROSAT observation. In order to search for spectral
variability we used the models derived from the XMM data
(Table\,\ref{br0351_spec}), folded those by the response matrix of the ROSAT
PSPC, and calculated the hardness ratios in the PSPC energy band.
For a powerlaw model with \ax=0.75 and \nh=4.08$\times~10^{20}$\cm~we derived a
HR=+0.51 and for \ax=1.19 and \nh=13.59$\times~10^{20}$\cm a HR=+0.93.
Although
this is a
rough estimate, it suggests that the source has become harder in the XMM
observation compared to the ROSAT PSPC observation 10 years before (in the
observed frame).

Note that a broken power-law model with $\alpha_{\rm X,soft}$=3.5 and
only Galactic absorption (Table \ref{br0351_spec}) fits the PN data
well, allowing consistency with the ROSAT data. On the other hand, the
model with excess absorption brings the expected count rate down by a
factor of 3.5, compared to the observed decrease by a factor of 6.
This means that a variable absorber can be responsible for most of the flux
variability observed in BR 0351--1034, but the X-ray source still has to be
intrinsically variable by a factor of about 2.

\subsubsection{Spectral Energy Distribution}

From the UV spectrum given in \citet{sto96b} we derived the optical
flux density at 2500\AA~ rest-frame. The rest-frame flux density at 2
keV was calculated based on the simple powerlaw fit to the EPIC PN
data (Table \ref{br0351_spec}).  This results in an optical to X-ray
spectral slope \aox=1.51. The 0.2-2.0 and 2.0-10 keV rest-frame X-ray
luminosities are log $L_{\rm 0.2-2.0 keV}$=45.5 [ergs s$^{-1}$] and
log $L_{\rm 2.0-10.0 keV}$=46.4 [erg s$^{-1}$]. This converts to a
bolometric luminosity of log $L_{\rm bol}~\approx~$46.7 ergs s$^{-1}$
using the relation\footnote{ $log~L_{\rm
bol}~=~0.44~+~1.02~\times~log~L_{\rm 0.2-2.0~keV}$. Note, that this
relation is determined for a Friedman cosmology with $H_0$=75 \kms
Mpc$^{-1}$ and $q_0$=0.0 and luminosities given in units of Watts.} 
given in \citet{gru03} derived for a low-redshifted soft X-ray
selected AGN sample.

\section{\label{discuss} Discussion}

\subsection{\label{dis_rxj1028} The radio-loud QSO RX J1028.6--0844}

As discussed above, RX J1028.6--0844 is a flat spectrum radio-loud
object. The high radio and X-ray luminosities suggests that the
emission is beamed and the source is possibly a BL Lac object, in
which we are viewing the source down the radio jet axis. The flat
X-ray spectral slope of this source is also consistent with this
interpretation. Also, the values of \aox~and \aro~ of RX J1028--0844
when placed in the \aox~vs. \aro~diagrams of \citet{lau99} and
\citet{bec03} argue for the BL Lac nature of this source.  However,
the presence of a Ly$\alpha$ emission line in the identification
spectrum of RX J1028--0844 \citep{zic97} and the lack of significant
X-ray variability in the source argue against a blazar classification.
Even though a lack of variability is unusual for a BL Lac object, it
is not unknown (e.g. \citet{tag03}).

Based on ASCA data, RX J1028.6--0844 was believed to have strong
intrinsic absorption (\S\,\ref{asca_comp}).  
Another high redshift blazar, PMN J0525--3344
also shows extremely strong X-ray absorption, again in ASCA spectrum,
more than a million times greater than the neutral hydrogen or dust
column density implied by its optical spectrum \citep{fab01}. Faced
with the difficulty of explaining this observation, \citet{fab01}
invoke highly ionized, and dust destroyed, absorbing gas close to the
nucleus, similar to the warm absorber commonly seen in Seyfert
galaxies.  Another z$>4$ blazar, GB 1428$+$4217, was also claimed to
have strong intrinsic absorption based on ASCA data
\citep{fab98}. Now our XMM-Newton observations question these
findings, because at least in the case of RX J1028.6--0844 we find the
ASCA results to be spurious. As shown in Figure\,\ref{rxj1028_pn_wapo}
and Table\,\ref{rxj1028_spec} the data are consistent with only
Galactic absorption. If present at all, the column density of the
intrinsic absorber is more than a factor of 10 smaller than that
previously suggested from ASCA observations by \citet{yua00}. The
superior sensitivity of the XMM EPIC PN detector down to energies of
0.2 keV \citep{str01} allows us to put much better constrains to the
models than with data of previous X-ray missions.

Our results are in accord with those of \citet{fer03} and
\citet{elv00}, who also found that the evidence for excess absorption
in PKS 2149--306 to be marginal based on XMM-Newton and BeppoSAX data
respectively, in contrast to previous ASCA results \citep{cap97}. It
is also interesting to note that ROSAT observations of GB 1428+4217,
which are similarly sensitive to low energies, also imply an order of
magnitude lower column density than the ASCA results
\citep{bol00}. The only radio-loud high redshift
quasar in which XMM observations confirmed the ASCA claims of excess
absorption is PKS 2126--158 \citep{fer03}. Note, however, that PKS
2126--158 is a giga-hertz peak spectrum (GPS) source \citep{dev97,
sta98}.  GPS sources are thought to be young radio-loud quasars
and/or in denser environments \citep{odea91}. The excess absorption
observed towards PKS 2126--158 may then be in its immediate
environment.

However, the claims of excess absorption towards high redshift
radio-loud quasars are not all from ASCA observations. \citet{elv94}
and \citet{cap97} obtained a similar result based on ROSAT
observations.  It should be noted though that the ROSAT results are
not very robust, with excess absorption in candidate radio-loud
sources with low-energy cutoffs being consistent with zero within
$2\sigma$ \citep{fiore98}. Better quality spectra with XMM-Newton
will certainly help resolve this issue, and the effort is already
underway.

The excess absorption towards RX J1028.6--0844 is weak, if present at
all, and as a result we can neither constrain its metalicity nor
redshift. \citet{yua00} have discussed in detail the possibility of
intervening material as the absorber and suggested a galaxy about
7$^{''}$ away from the position of RX J1028.6--0844 found in the image
obtained by \citet{zic97} as a possible site. \citet{per01} found a
weak damped Ly$\alpha$ system at z=3.42 and z=4.05 and estimated the
column density to be log $N_{HI}$ $^{>}_{\sim}$ 20.1 [\cm]. We fitted
the XMM-Newton PN and MOS data simultaneously to a redshifted absorber
with z=3.42 and solar abundance resulting in an absorption column
$N_H$=0.93$\times$ 10$^{22}$\cm. If the abundance is sub-solar,
similar to that found in other damped Ly$\alpha$ systems
\citep{bec01}, the effective total column density would be even
higher. The z=3.42 damped system is thus an unlikely site for the
X-ray absorption, but a lower redshift damped system is a
possibility. It is also quite possible that the excess absorption, if
present, is local to our own Galaxy. The observed Galactic column
density is averaged over 1 square degree based on the HI maps of
\citet{dic90} and it is quite possible that the actual column density
is somewhat higher either due to a neutral or molecular cloud.

The major objective of our observing program was to compare the
properties of high redshift quasars to their low redshift cousins. For
this reason, we placed RX J1028.6--0844 on the $\alpha_{\rm ro}$
vs. $\alpha_{\rm ox}$ plot for complete samples of low redshift
radio-loud quasars (\citet{fab99}; please note that \citet{fab99}
define their spectral energy slopes at 5500 \AA\ and 1keV instead of
conventional 2500 \AA\ and 2keV). We find our source to lie right in
the region occupied by the BL Lac objects. Comparison with a similar
plot by \citet{sie98}, where BL Lacs occupy the regions of
0.6$<$\aox$<$1.6 and 0.25$<\alpha_{\rm ro}<$0.75, also shows that RX
J1028.6--0844 does not occupy any conspicuously different region.
Thus the present observation does not show any evolution of broad band
properties with redshift.  The \aox~ value of RX J1028.6--0844
occupies lower end of the distribution of radio-loud quasars by
\citet{wil94}, implying larger X-ray luminosity, if any, compared to
the lower redshift quasars.

\subsection{The radio-quiet quasar BR0351--1034}


BR 0351--1034 is variable by a factor of about 6 on a timescale of
1.75 years (in the rest-frame). While variabilities in X-rays with
factors of 3 or 4 on timescales of days are quite common among AGN
(e.g. \citet{lei99, gru01}), evidence of strong variability among high
redshift quasars is sparse. With the present spectral quality, we
cannot distinguish among following scenarios: (1) flux of BR
0351--1034 decreased from the time of ROSAT observation to the time of
XMM-Newton observations, but the spectrum did not change; (2) flux
decreased in part due to excess absorption during the XMM-Newton
observation; and (3) flux decreased and the spectrum hardened.  All
the three types of variations have been observed in low redshift AGN:
e.g. in IRAS1334+24 the flux changed between the ROSAT All Sky Survey
and the pointed observation by over a factor of two, but the spectral
shape did not change \citep{gru01}; in NGC 3516 spectral change due
to change in absorber column density has been reported \citep{mat97}
and in the Narrow-Line Seyfert 1 galaxy RX J0134.2--4258
\citep{gru00} spectral change is observed with change in flux.

The UV/optical - X-ray broad-band spectral slope of BR0351--1024 is
\aox=1.51. This is as expected for a radio-quiet quasar with optical
luminosity density of log $l_{\rm o}$=31.7 ergs s$^{-1}$ Hz$^{-1}$
(\citet{yua98}, see their figure 11). Thus, our observations do not
support earlier claims that high redshift radio-quiet quasars are
weaker in X-rays compared to lower-z quasars (\citet{bri97} and
\citet{fer03}).

The most interesting, but tentative result is that BR0351--1024 has a
very steep soft X-ray slope. In order to reconstruct the ROSAT PSPC
hardness ratio HR=--0.24 a steep soft X-ray slope \ax=3.5 is needed
and the XMM-Newton spectrum is also consistent with such a steep slope
(Sect.\,\ref{res_xmm}).  While such steep X-ray spectra have been
observed in a number of Narrow-Line Seyfert 1 galaxies
(e.g. \citet{bol96, gru98, gru01}), only one other high redshift
quasar (SDSS J1044-0125 at z=5.8) with possible steep X-ray spectrum
is known 
(\citet{mat01}; note that the 0.1--0.4 keV range in the
observed soft band corresponds to rest frame 0.53--2.1 keV in the rest
frame of BR0351--1034, similar to the ROSAT band). It is commonly
believed that the steep soft X-ray spectra in NLS1s are the result of
close to Eddington accretion (e.g. \citet{pou95}). Is BR0351--1034
also in a very high accretion state? Given the bolometric luminosity
of log $L_{\rm bol}~\approx~$46.7 [ergs s$^{-1}$] Eddington accretion
rate implies a black hole mass of about a few times $10^8$\msun.  
According to
accretion-disk corona models (e.g. \citet{kur00}), it is difficult to
generate such steep X-ray spectra around such massive black holes. Or
is the soft X-ray emission a result of Compton thick outflows
associated with Eddington or super-Eddington sources \citep{king03}? 
These are intriguing questions, which we can answer only after
confirming the suggestive evidence presented here. If true, the
steepness of the soft X-ray slope in BR0351--1034 supports the
proposal of \citet{mat00} that high redshift quasars and NLS1s are
similar objects, perhaps at an early stage of their evolution
(\citet{gru96, gru99} and \citet{mat00}). This hypothesis has been
recently supported by the findings of \citet{yuan03} who showed that
high redshift quasars are at the same extreme end of the \citet{bor92}
'Eigenvector 1', as the NLS1s.


\section{Summary \& Conclusion}

We have studied the XMM-Newton data of the high redshift blazar RX
J1028.6--0844 and a radio-quiet quasar BR0351--1034. We found that the
evidence of excess absorption towards RX J1028.6--0844 is weak at
best. If present, the column density of the redshifted absorber is
more than 20 times smaller than what has been previously suggested
from ASCA data.  Location of the absorber is unconstrained, and it may
well be in our Galaxy.  A longer, 40 ks XMM-Newton observation in AO2
(PI W. Yuan) will be valuable to confirm the excess absorption if any
and to get better constrains on the location and metallicity of the
absorber. Our observations of BR0351--1034 were compromised due to
high radiation background, but the present data do not support claims
of X-ray weakness in high redshift radio quiet quasars. Similarly, the
X-ray properties of RX J1028.6--0844 do not appear to be significantly
different from low redshift BL Lac objects. Clearly, we cannot draw
any definite conclusions about quasar evolution from just two
observations, and we will publish the results from our entire sample
as and when all the observations are made. There is a tantalizing
evidence of steep soft X-ray slope in this source, supporting the
hypothesis of \citet{mat00} about the evolution of AGN. We also have
more XMM-Newton observations of radio-quiet quasars approved in cycle
2 and 3, including a longer observation of BR0351--1034, which will
help confirm and extend the results presented here.

\acknowledgments

We would like to thank Matthias Dietrich for the discussions on
abundances in high redshift quasars, Weimin Yuan for intensive discussions on
the ASCA observation of RX J1028--0844, Michael Freyberg for
discussions on calibration issues, David Weinberg for providing a program to
determine the luminosity distances of the sources, and the anonymous referee for
a fast referee's report and useful suggestions on the manuscript.
This research has made use of the
NASA/IPAC Extra-galactic Database (NED) which is operated by the Jet
Propulsion Laboratory, Caltech, under contract with the National
Aeronautics and Space Administration, and data from the US Naval Observatory
catalogue A2.0.
The ROSAT project is supported
by the Bundesministerium f\"ur Bildung und Forschung (BMBF/DLR) and
the Max-Planck-Society. 
 This work was supported in part by NASA grant
NAG5-9937.

\clearpage


\begin{figure*}
\epsscale{1.7}
\chartlineb{DGrupe.fig1a}{DGrupe.fig1b}
\caption{\label{q1028_plot} Power-law fit with neutral galactic absorption
(fixed to galactic value) and intrinsic absorption with metal
abundance = solar to the EPIC PN, MOS-1 and 2, and ASCA SIS0 and SIS 1
data of RX J1028.6--0844 (\ax=0.272, $N_{\rm H, intr}~=~0.79\times
10^{22}$\cm, Table\,\ref{rxj1028_spec_zvfeabs}).  The left panel shows
the fits to the spectra of the EPIC PN (top) MOS (middle), and ASCA
SIS (bottom) detectors.  The right panel displays the 68\%, 90\% and
99\% confidence levels of the Column density of the intrinsic absorber
vs. the Photon index (Photon index $\Gamma$=\ax+1;
Tab\,\ref{rxj1028_spec_zvfeabs}).}
\end{figure*}

\begin{figure}
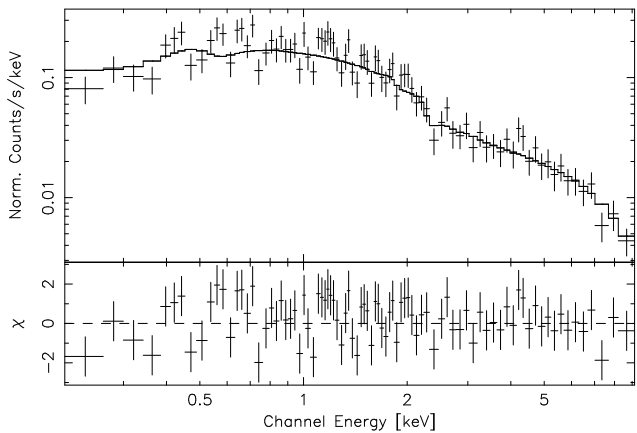

\epsscale{1.2}
\clipfig{DGrupe.fig2}{87}{5}{5}{250}{167}
\caption{\label{rxj1028_pn_wapo} Powerlaw model with Galactic absorption 
 fitted to the PN data of RX J1028.6--0844 (\ax=0.233, $N_{\rm
 H}~=~4.59\times 10^{20}$\cm; Table\,\ref{rxj1028_spec}). }
\end{figure}

\begin{figure}
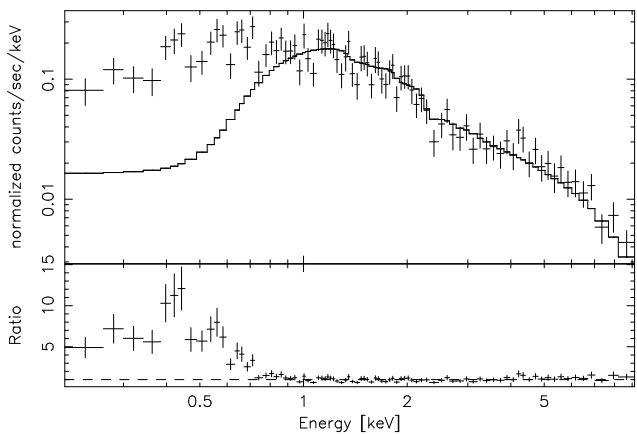

\epsscale{1.2}
\clipfig{DGrupe.fig3}{87}{5}{5}{250}{167}
\caption{\label{rxj1028_pn_ascapara} Power-law plus neutral Galactic and
redshifted intrinsic absorption using the values from the fit to the
ASCA data of RX J1028--0844 (\citep{yua00}, $N_{\rm H,
intr}~=21.1\times10^{22}$\cm, \ax=0.72).  The figure clearly shows
that the XMM data agree with the ASCA model for energies $>$ 1keV but
deviate significantly in the soft X-ray band.  }
\end{figure}



\begin{figure}
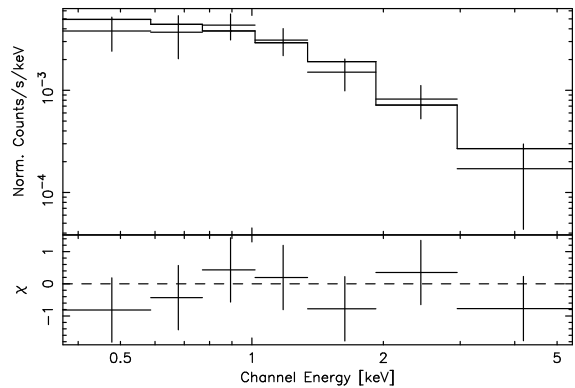

\clipfig{DGrupe.fig5}{87}{1}{5}{275}{170}
\caption{\label{br0351_specfits_galnh} Power-law model with neutral Galactic 
absorption fitted to the EPIC PN data of BR 0351--1034 (\ax=0.75,
$N_{\rm H}~=~4.08\times 10^{20}$\cm; Table\,\ref{br0351_spec}). .  }
\end{figure}

\begin{figure}
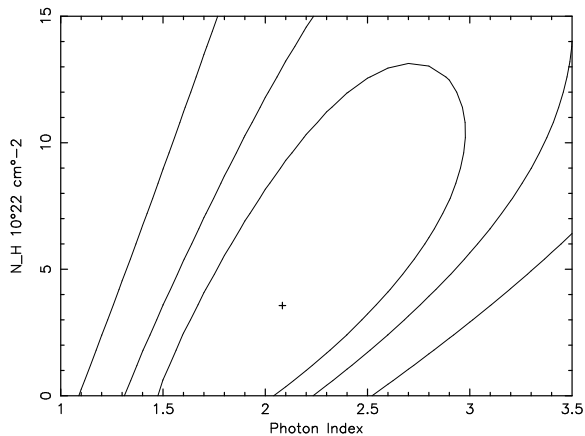

\clipfig{DGrupe.fig6}{87}{2}{0}{275}{185}
\caption{\label{br0351_speccontour} Contour plot with
 the 68\%, 90\% and 99\% confidence levels of the Column density of
 the intrinsic absorber vs. the Photon index of a powerlaw model with
 Galactic and intrinsic neutral absorption fit to the EPIC PN data of
 BR 0351--1034 (Table\,\ref{br0351_spec}). }
\end{figure}




\begin{deluxetable}{lccr}
\tabletypesize{\scriptsize}
\tablecaption{Spectral Fit parameters of RX J1028.6--0844 for a power-law
model with neutral absorption at z=0. 
 \label{rxj1028_spec}}
\tablewidth{0pt}
\tablehead{
&  \colhead{$N_{\rm H}$} & \\
\colhead{\rb{Detector}} & \colhead{10$^{22}$\cm}     &
\colhead{\ax} & \colhead{\rb{$\chi^2$ (DOF)}} 
}
\startdata
PN\tablenotemark{1} & 0.0665\pl0.0128 & 0.307\pl0.054 & 91.8 (84) \\
& 0.0459 (fixed) & 0.233\pl0.032 & 95.2 (85) \\
MOS-1\tablenotemark{2} & 0.1653\pl0.0486 & 0.424\pl0.110 & 40.3 (35) \\
& 0.0459 (fixed) & 0.194\pl0.595 & 47.0 (36) \\
MOS-2\tablenotemark{2} & 0.0883\pl0.0451 & 0.314\pl0.112 & 31.5 (37) \\
& 0.0459 (fixed) & 0.225\pl0.060 & 32.4 (38) \\
MOS-1\tablenotemark{2} \& MOS-2\tablenotemark{2} & 
0.1251\pl0.0332 & 0.368\pl0.079 & 73.0 (74) \\ 
& 0.0459 (fixed) & 0.209\pl0.042 & 79.4 (75) \\
ASCA SIS-0\tablenotemark{3} \& SIS-1\tablenotemark{4}  & 
0.0611\pl0.1465 & 0.416\pl0.207 & 26.3 (28) \\
& 0.0459 (fixed) & 0.397\pl0.100 & 26.5 (29) 
\\ \\
PN\tablenotemark{1} + MOS-1\tablenotemark{2} \& MOS-2\tablenotemark{2} & 
0.0712\pl0.0119 & 0.298\pl0.042 & 169.0 (160) \\
& 0.0459 (fixed) & 0.225\pl0.025 & 174.6 (161) \\
PN\tablenotemark{1} + ASCA SIS-0\tablenotemark{3} \& SIS-1\tablenotemark{4}  & 
0.0715\pl0.0122 & 0.331\pl0.047 & 119.0 (113) \\
& 0.0459 (fixed) & 0.249\pl0.031 & 124.5 (114) \\
PN\tablenotemark{1} + MOS-1\tablenotemark{2} \& MOS-2\tablenotemark{2} + 
ASCA SIS-0\tablenotemark{3} \& SIS-1\tablenotemark{4} & 
0.0725\pl0.0114 & 0.304\pl0.040 & 211.8 (189) \\
& 0.0459 (fixed) & 0.229\pl0.025 & 206.0 (190) 
\enddata

\tablenotetext{1}{Energy range 0.2-10 keV}

\tablenotetext{2}{Energy range 0.5-7.5 keV}

\tablenotetext{3}{Energy range 0.8-7.0 keV}

\tablenotetext{4}{Energy range 0.8-6.5 keV}
\end{deluxetable}

\begin{deluxetable}{lcccr}
\tabletypesize{\scriptsize}
\tablecaption{Spectral Fit parameters of RX J1028.6--0844 for a power-law
model with galactic neutral absorption at z=0 and 
redshifted intrinsic absorption.
The galactic absorption is fixed to $N_{\rm H, gal}=4.59~10^{20}$ \cm~
(\citep{dic90}).
 \label{rxj1028_spec_zvfeabs}}
\tablewidth{0pt}
\tablehead{
&  \colhead{$N_{\rm H, intr}$} &  \\
\colhead{\rb{Detector}} & \colhead{10$^{22}$\cm}  &  
\colhead{\ax} & \colhead{\rb{$\chi^2$ (DOF)}} 
}
\startdata
PN\tablenotemark{1} &  0.781\pl0.447  & 0.300\pl0.048 & 91.0  (84)\\
MOS-1\tablenotemark{2} & 5.57\pl2.68  & 0.378\pl0.105 & 41.7  (35) \\
MOS-2\tablenotemark{2} & 2.17\pl2.38  & 0.302\pl0.203 & 31.4  (37) \\
MOS-1\tablenotemark{2} \& MOS-2\tablenotemark{2} & 
4.07\pl1.83  &  0.347\pl0.075 & 78.2 (75) \\
ASCA SIS-0\tablenotemark{3} \& SIS-1\tablenotemark{4}  & 
17.90\pl14.16  & 0.748\pl0.321 & 24.8 (28) \\
PN\tablenotemark{1} + MOS-1\tablenotemark{2} \& MOS-2\tablenotemark{3} & 
0.87\pl0.41  & 0.277\pl0.036 & 169.7 (160) \\
PN\tablenotemark{1} + MOS-1\tablenotemark{2} \& MOS-2\tablenotemark{2}
 +  ASCA SIS-0\tablenotemark{3} \& SIS-1\tablenotemark{4} & 
0.89\pl0.43 & 0.280\pl0.035 & 200.7 (189) 
\enddata

\tablenotetext{1}{Energy range 0.2-10 keV}

\tablenotetext{2}{Energy range 0.5-7.5 keV}

\tablenotetext{3}{Energy range 0.8-7.0 keV}

\tablenotetext{4}{Energy range 0.8-6.5 keV}

\end{deluxetable}

\begin{deluxetable}{lccccr}
\tabletypesize{\scriptsize}
\tablecaption{Spectral Fit parameters to the EPIC PN data of BR 0351--1034
 \label{br0351_spec}}
\tablewidth{0pt}
\tablehead{
&   \colhead{$N_{\rm H, gal}$} & \colhead{$N_{\rm H, intr}$} \\
\colhead{\rb{XSPEC Model}} & 
\colhead{10$^{20}$\cm}  & \colhead{10$^{22}$\cm}   
& \colhead{\rb{$\alpha_{\rm X, soft}$}} &
\colhead{\rb{$\alpha_{\rm X, hard}$}} &  \colhead{\rb{$\chi^2$ (DOF)}} 
}
\startdata
wa powerlaw\tablenotemark{1}  & 13.59\pl12.79 & --- & --- &
1.19\pl0.66 & 1.2 (4) \\
 & 4.08 (fixed)  & --- & --- & 0.75\pl0.27 & 2.4 (5) \\
wa zwa po\tablenotemark{2}  & 4.08 (fixed) & 3.58\pl4.45 & --- &
1.08\pl0.48 & 1.1 (4) \\
wa bknpo\tablenotemark{3} & 4.08 (fixed) & --- & 3.50 (fixed)
& 0.73\pl0.28 & 3.0 (4) \\
& 12.70\pl11.81 & --- & 3.50 (fixed) & 1.13\pl0.62 & 1.2 (3)
\enddata

\tablenotetext{1}{galactic absorption and powerlaw model}

\tablenotetext{2}{galactic absorption, redshifted neutral absorption at z=4.351
and powerlaw}

\tablenotetext{3}{galactic absorption
and broken powerlaw with observed break energy $E_{\rm break}$=0.45\pl0.18 keV.
}
\end{deluxetable}

\end{document}